\documentclass[12pt]{article}
\usepackage{times}
\usepackage{graphicx}
\usepackage {amsfonts}
\usepackage {amssymb}
\usepackage {namedplus}
\begin{document}
\newfont{\fntbig}{cmssbx10 scaled 1728}
\newfont{\fntmed}{cmssbx10 scaled 1200}
\newfont{\fntsm}{cmssbx10}
\newfont{\fntplace}{cmss9}
\newfont{\fntcap}{cmss8}
\newfont{\fntrefs}{cmr7}
\newfont{\fntrefsi}{cmsl7}
\newfont{\fntrefsb}{cmbx7}

\noindent
{\fntbig Localization and extinction of bacterial populations
under inhomogeneous growth conditions}\\
\vspace{0.1truein}

\noindent
{\fntmed
Anna~L.~Lin$^{*\dagger}$, Bernward~A.~Mann, Gelsy~Torres-Oviedo, 
Bryan~Lincoln, Josef~K$\bf{\ddot{a}}$s,\\
Harry~L.~Swinney}\\
\noindent
{\fntplace
Center for Nonlinear Dynamics and Department of Physics,
The University of Texas at Austin, Austin, TX 78712}\\
{\fntplace
$^*$To whom correspondence should be addressed. email: alin@phy.duke.edu\\
$^\dagger$Current address:
Center for Nonlinear and Complex Systems and Department of Physics,
Duke University, Durham, NC 27708}

\vspace{0.1truein} {\fntsm \noindent The transition from localized
to systemic spreading of bacteria, viruses and other agents is a
fundamental problem that spans medicine, ecology, biology and
agriculture science. We have conducted experiments and simulations
in a simple one-dimensional system to determine the spreading 
of bacterial populations that occurs for an inhomogeneous environment
under the influence of external convection.
Our system consists of a long channel with growth inhibited by
uniform UV illumination except in a small ``oasis'', which is
shielded from the UV light. To mimic blood flow or other flow past
a localized infection, the oasis is moved with a constant velocity
through the UV-illuminated ``desert''. The experiments are modeled 
with a convective reaction-diffusion equation.  In
both the experiment and model, localized or extinct populations
are found to develop, depending on conditions,
from an initially localized population.
The model also yields states where the
population grows everywhere. Further, the model reveals that the
transitions between localized, extended, and extinct states are
continuous and non-hysteretic. However, it does not capture
the oscillations of the localized population that are observed in
the experiment.} \vspace{0.1truein}

\renewcommand{\baselinestretch}{1}\large\normalsize  

\noindent The growth, spreading, and extinction of a population in
an inhomogeneous environment is of interest given the global
decline in biodiversity.  Under what conditions does a population
change from one that is localized to one that spreads throughout
the available domain?  This question is important 
not only for species in the earth's ecosystem but also 
for understanding the transition from a localized to a
systemic infection in an organism. Simple model systems can
provide insight into the kinds of phenomena that can occur in
these complex systems.  Population growth studies often examine
bacteria because of the short doubling time.  Studies of bacterial
growth in homogeneous environments have revealed the spontaneous
formation of different spatial patterns in response to
environmental stresses \cite{ShDw:97,BePu:77,BSTC:94,MWIR:98}. We
are interested in the situation more commonly found in nature:
inhomogeneous environments.

Our work is motivated by recent theoretical studies of 
the Fisher equation, which was generalized to model
growth in an inhomogeneous environment under the influence of a
convective flow. Those studies indicated a
transition from localized populations to delocalized populations
that grow everywhere \cite{NeSh:98,DNSh:00,Shne:01}. Experiments
have also been conducted on growth in inhomogeneous environments,
but the conditions were too complex to
compare with the model studies \cite{NPLK:00,Shne:01}.

We have developed an experimental system for studying bacterial
growth in an inhomogeneous environment. $E.\ coli$ bacteria
grow in a long, thin channel with different growth conditions in a
small section of it.  The position of this different growth region
changes in time.  The quasi-one-dimensional geometry permits
comparison with simple one-dimensional models. We measure the
bacterial concentration as a function of space and time in
experiments that typically last one week, which is long compared
to the half-hour doubling time for the bacteria. We integrate the
full nonlinear partial differential equation model. 

\vspace{0.1truein}

\noindent

The spreading of bacterial colonies measured for certain
homogeneous conditions \cite{ShDw:97,MWIR:98} in a petri dish has
been found to be described by the Fisher equation 
\cite{Fish:37,Murr:89}, which has
traveling front solutions that propagate with a velocity $v_F =
2(Da)^{1/2}$, where $a$ and $D$ are respectively the growth rate
and diffusion coefficient of the bacteria
\cite{ShDw:97,BePu:77,BSTC:94}.  Nelson et al.
\cite{NeSh:98,DNSh:00} proposed a generalization of the Fisher
equation to describe the growth of bacteria in an environment with
a localized region of favorable growth conditions (an oasis) that
moves with a constant velocity $v$:

\begin{equation}
\label{gfisher} \frac{\partial c(x,t)}{\partial t} = D \nabla^2
c(x,t) + U(x- vt)c(x,t) -  b c(x,t)^2.
\end{equation}
\begin{figure}
\centering
\includegraphics[width=3.3truein]{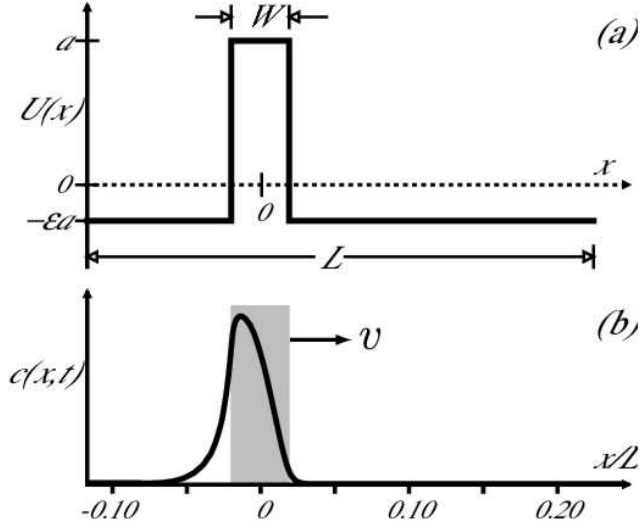}\\
{\renewcommand{\baselinestretch}{1}
\caption{\label{schematic} (a) Growth profile for a system
with an oasis (width $W$ and growth rate $a$) that moves with
velocity $v$ through a desert (width $L-W$ and growth rate
-$\epsilon a$). (b) Example of a localized colony after 69 hours
of growth, from a numerical simulation of Eq.~\ref{gfisher}. The
colony density is shown by the solid black line and the gray band
depicts the oasis. Simulation parameter values: $v/v_F = 0.9$, $a
= 4.76 \times 10^{-4}$ s$^{-1}$, $D = 6 \times 10^{-5}$ cm$^2/$s,
$\epsilon = 0.7$, $W = 2$ cm, $L = 50$ cm, and 
$b = 7.93 \times 10^{-11}$ cm/s. 
Concentration $c(x,t)$ is in arbitrary units.}}
\end{figure}
The growth rate in the oasis ($|x| \leq W/2$)  
is $U(x - vt) = a$, 
and in the surrounding desert 
($|x| > W/2$) the growth rate is $U(x - vt) = -\epsilon a$, 
where $W$ is the width of the oasis in a system of length $L$, as
shown in Fig. ~\ref{schematic}(a). 
The average growth rate over the full length of the system is

\begin{equation}
\label{terrain} \frac{\langle U(x - vt) \rangle}{a} = \frac{W
-\epsilon(L-W)}{L},
\end{equation}
which we call the average terrain. An analysis of a linearization of 
Eq.~\ref{gfisher} by Nelson et al. \cite{NeSh:98,DNSh:00}
suggested that qualitatively different spatial distributions of
concentration would occur for different values of the oasis
velocity $v$ and the average terrain (Eq.~\ref{terrain}).
Numerical simulations \cite{DNSh:00} of the full 
nonlinear equation suggested
that these qualitative changes occur not over a range of parameter
values, but rather at particular thresholds of these two
parameters; however, this was not investigated in detail.

\vspace{0.1truein}

\noindent {\fntsm Extinction, localization, and delocalization.}
For oasis velocities $v$ exceeding a critical value $v_c$ and
sufficiently harsh terrains (i.e., for $\frac{\langle U
\rangle}{a}$ sufficiently negative), life is not sustainable and
the population becomes extinct, approaching zero concentration
everywhere.

For $|v| < v_c$, there is a range of values of the terrain
$\frac{\langle U \rangle}{a}$ for which the population remains
localized, growing in the oasis while decaying exponentially with
distance from the oasis. An example of such a localized state is
pictured in Fig.~\ref{schematic}(b). The population is sustained
in the moving oasis but dies out in the desert.

Finally, for a sufficiently rich terrain the population grows
everywhere, for any value of the oasis velocity $v$. No matter how
the population is initially distributed throughout the length of
the system, the population becomes delocalized.

The identification of the distinct sustainable localized,
delocalized, and extinct states, and the suggestion that there
could be well-defined transitions among them, originally resulted
from an analysis of a linearized version of Eq.~\ref{gfisher},
which disregards the biologically important saturation term
\cite{NeSh:98,DNSh:00}. Transitions between the different regimes
can be understood only by including the nonlinear saturation term.
We have performed numerical simulations of the full nonlinear
equation to compare with the laboratory observations and to study
the transitions among the localized, delocalized, and extinct
states. \vspace{0.1truein}

\noindent
{\fntmed Methods}\\
We examined bacterial growth in a rectangular channel of length
$L$ =25 cm, thickness 0.2 cm, and height 2.5 cm. This thin
quasi-one-dimensional geometry was chosen so that the observations
could be compared to predictions of one-dimensional models
\cite{DNSh:00}.  Most of the 25 cm long channel was exposed to UV
light to inhibit bacterial growth, but inside this "desert" was a
short ($W=3$ cm) "oasis" where the UV light was blocked by a mask
(see Fig. 1(a)).

The species selected for this study was $E.\ coli$ RW 120
because, for these bacteria,
UV exposure results in cell death rather than mutation (the two
mechanisms for repair of DNA are switched off) \cite{source}.
Thus, as in the model, the population dies out in the desert if it
is sufficiently harsh.

The rectangular channel was filled before each experimental run
with a define media in a 0.2\% agar gel maintained at 37 $^\circ$C;
this nutrient level supported steady bacterial growth for at least
one week. The purpose of the gel was to inhibit uncontrolled
convection while
allowing diffusion of the bacteria. A 120 cm long mercury lamp
positioned 8.5 cm above the cell provided 10 W/m$^2$ UV
illumination for the region not shielded by the mask.

Each run was initiated by dragging a needle that had been dipped
into bacteria across the gel surface underneath the mask, thus
infecting only the oasis region. We waited several hours for the
bacteria to reach a saturated concentration $c_s=a/b$ in the oasis 
before starting to move the oasis with a constant velocity
$v$, as in Fig. 1(b). Runs were made for oasis velocities ranging
from 1.5$\times 10^{-5}$ cm/s, where the oasis moved only 13 cm in an
experiment lasting ten days, to 3.6$\times 10^{-4}$ cm/s, where
the oasis traversed the 25 cm cell length in 19 h. At high oasis
velocities, the original population could not keep up with the
moving oasis, and the population decreased and became extinct; we
continued to monitor the decline in the original population for
several days after the oasis reached the end of the cell. During a
week-long experiment, most of the live bacteria remained close to
the (oxygenated) air-gel interface, while dead bacteria sunk to
the bottom of the channel in about two days.

The bacterial concentration near the gel surface was determined by
measuring the transmission of light through the thin dimension of
the cell. The light source was a 4.5 mW diode laser with a 670 nm
wavelength and a 1 mm beam diameter. The light source and a
detector were mounted on a translation stage with the center
of the beam passing
through the gel 1 mm below its surface. Each spatial scan of the
bacterial concentration took 7 minutes, a time short compared to
the time for the concentration to change significantly; thus the
measurements yielded $c(x,t)$ at all positions $x$ and times $t$
for the duration of an experiment (3-10 days). \vspace{0.1truein}

\noindent {\fntsm Experimental determination of the model
parameters.}
We have measured the parameters in the model: the 
growth rate $a$ of $E.\ coli$ in 0.2\% agar define media, the growth
attenuation coefficient $\epsilon$ of the UV-light, the diffusion
constant $D$, and the saturation constant $b$.

\begin{figure}
\centering
\includegraphics[width=4.1truein]{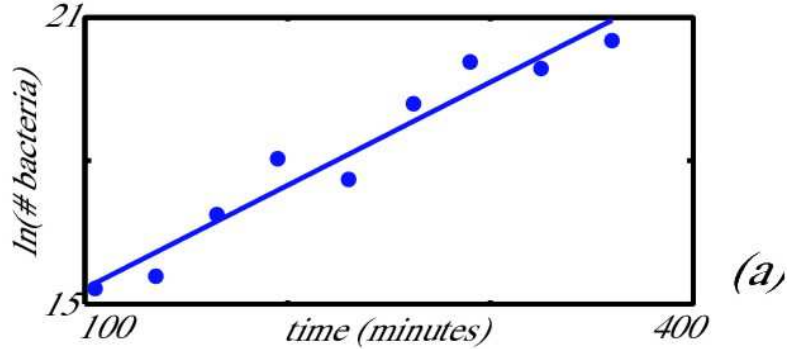}
\vspace{-0.1truein}
\caption{\label{parameters} {Experimental determination of
the growth rate $a$.  ("\#bacteria" is the number
of bacteria per 100 $\mu$ml.)}}
\end{figure}
The growth rate of RW 120 was determined using a standard dilution
plating technique \cite{laura} in which the number of bacteria per
unit volume was counted every half-hour. The measurements, shown
in Fig.~\ref{parameters}, yielded $a = 4 \times 10^{-4}$
s$^{-1}$.

To determine $\epsilon$ we measured the bacterial death
rate caused by the UV light and find $\epsilon = 28$;
hence $\frac{\langle U \rangle}{a} = -24.5$.

The diffusion constant $D$ was determined assuming that
$D=v_F^2/2a$ \cite{Murr:93}, where $v_F$ was taken to be the
radially averaged front velocity of a colony growing in 0.2\% agar
in a petri dish. For bacteria growing in a petri dish with 0.2\%
agar in define media, we measured $v_F = 2.4 \times 10^{-4}$ cm/s.
This result together with the measured value of $a$ yields $ D= 7
\times 10^{-5}$ cm$^2/$s.

A steady state concentration $c_s$ was achieved with point
inoculation of bacteria in the oasis and no incubation period for
$v/v_F$ $\ll$ 1. A value for the saturation parameter $b$, $7.93
\times 10^{-11}$ cm/s, was determined from the relationship $b
= a/c_{s}$.  

We also conducted some measurements in 0.1\% agar LB media where
$a$ and $D$ were larger. These measurements yielded $a = 6\times
10^{-4}$ s$^{-1}$, $v_F = 4.4 \times 10^{-4}$ cm/s, and $D = 1.6
\times 10^{-4}$ cm$^2/$s.

\vspace{0.1truein}

\noindent {\fntsm Simulation.} We numerically integrated
Eq.~\ref{gfisher} using the parameter values given in
the previous section, except for $\epsilon$ which was
scanned over a range of values in the simulations, 
see Fig.~\ref{phasedia}. 
The value of $v_F$ used 
in the simulations was typically $v_F = 3.38 \times 10^{-4}$
cm/s, the average of the values measured in the two different
growth media. In the simulation, space was discretized using the
method described in \cite{DNSh:00,FrNe:00}. The resulting $N$
coupled ordinary differential equations (where the number of grid
points $N$ was typically 2032) were integrated in time assuming as
initial condition a Gaussian concentration profile with 10\% noise
centered in the favorable growth region \cite{Mann:01}.  An
alternative initial condition, a spatially uniform concentration
with the saturation value $a/b$, also with 10\% noise, yielded the
same behavior at long times. Both periodic and reflective boundary
conditions were found to yield extinct and localized solutions,
while delocalized solutions were found only with periodic boundary
conditions. \vspace{0.1truein} 

\begin{figure}
\centering
\includegraphics[height=6.9truein]{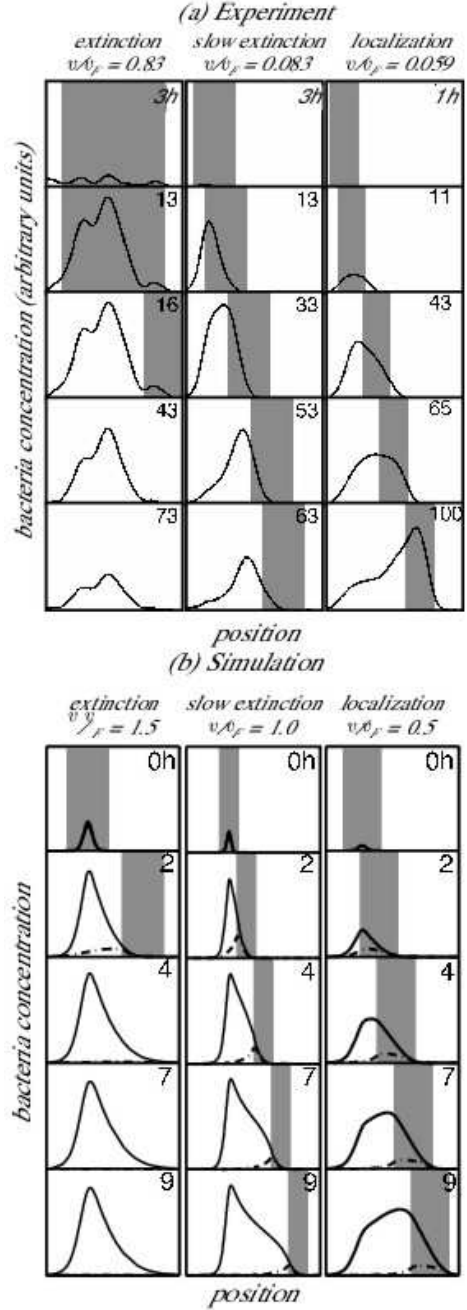}
\caption{\label{expsim} Bacterial colony evolution in (a)
experiment and (b) numerical simulation of Eq.~\ref{gfisher}, for
oasis velocities $v$ leading to extinction (left column), slow
extinction (middle column), and a sustained population in the
oasis (right column).  The gray boxes show the location of the
oasis (3 cm wide in a sample of length 25 cm); in the left column
the oasis moves out of the region shown, leaving behind the
population (which initially developed in a stationary oasis)
exposed to the deadly UV radiation. The same thing happens in the
middle column, but more slowly. The experimental measurements
cannot distinguish between live and dead bacteria, but the results
from from the simulation show the concentration $c(x,t)$ of
bacteria living at time $t$ as dashed lines, while solid lines
show the total concentration $\int c(x,t) dt$ of living and dead
bacteria; thus at $t=$ 2 h almost all of the bacteria are already
dead in the simulation (first column). Other simulation parameters
are given in \protect{Fig.~\ref{phasedia}. For the simulations the
bacterial concentration varies from 0 to 1.97$a/b$, 2.74$a/b$, and
10.73$a/b$ in the left, middle, and right columns, respectively.}}
\end{figure}

\noindent
{\fntmed Observation of localization and extinction}\\
\noindent {\fntsm Comparison of experiment and simulation.} In the
experiment we varied a single parameter, the velocity $v$ of the
oasis. The observations are compared with the numerical
simulations in Fig.~\ref{expsim}: for large $v$ the population
becomes extinct (left column) and for small $v$ a sustained
localized population develops (right column); the figure also
shows an example of an intermediate value of $v$, where the
population slowly becomes extinct (middle column).  The model
captures the observed qualitative dependence on $v$. 

In the remainder of this subsection we describe further the
experimental observations, and in the next subsection we present
results from the simulations, where an extensive exploration of
the control parameter space was possible, including the parameter
range yielding delocalized populations, which were not studied in
the experiments because of the long evolution times required and
because the current set-up has non-periodic boundary conditions.

In the case shown in Fig.~\ref{expsim}(a), the bacteria inoculated
in the oasis were first allowed to grow for 13~h with the oasis at
rest before it began to move at 3.6$\times 10^{-4}$ cm/s. The
bacteria did not diffuse and grow fast enough to maintain a 
population within the moving oasis, and the original bacterial
colony that was left behind became fully exposed to UV light. The
bacteria then died and slowly sunk to the bottom of the channel,
out of the field of the measurement of concentration; at 73 h some
(presumably dead) bacteria were still present, but the population
was declining toward zero. The waviness in the concentration
profile is due to the inhomogeneous inoculation of the gel (see
$t$=3 h).

The development of stable localized populations in a moving oasis
is illustrated in the column on the right of Fig.~\ref{expsim}.
The bacterial colony moves in the same direction as the oasis
motion, and at least part of the colony remains in the oasis;
bacteria that are left behind in the desert eventually die.

\begin{figure}
\centering
\includegraphics[width=4.25truein]{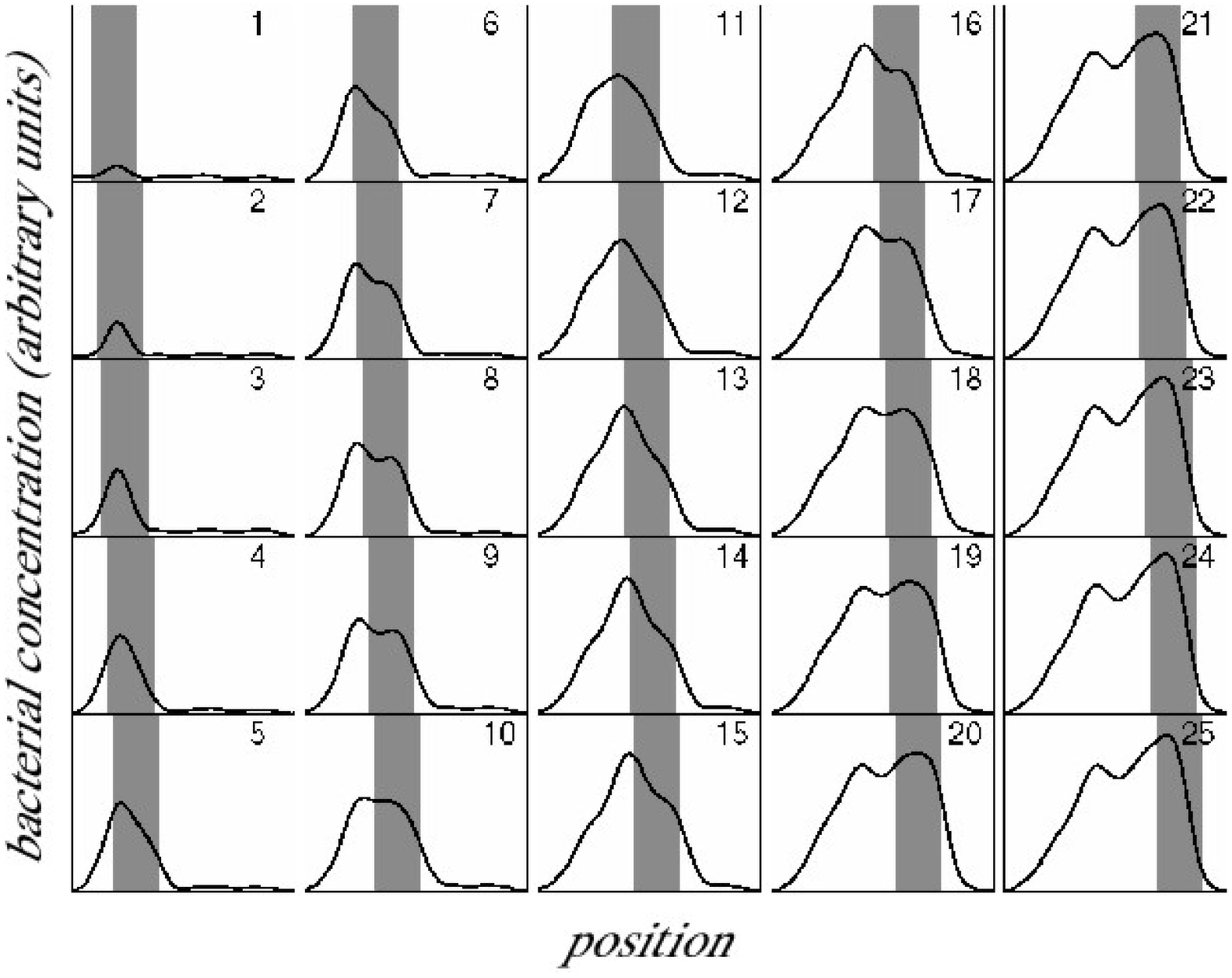}
\vspace{-0.1truein}
\caption{\label{oscillate} Snapshots taken hourly in the
experiments reveal spatio-temporal oscillations of the bacterial
concentration. The gray boxes show the oasis position. The
velocity of the oasis was $v$ = 1$\times$10$^{-4}$ cm/s, and the
experiment was conducted for a 0.1\% agar gel with LB media.}
\end{figure}

\begin{figure}
\centering
\includegraphics[width=4truein]{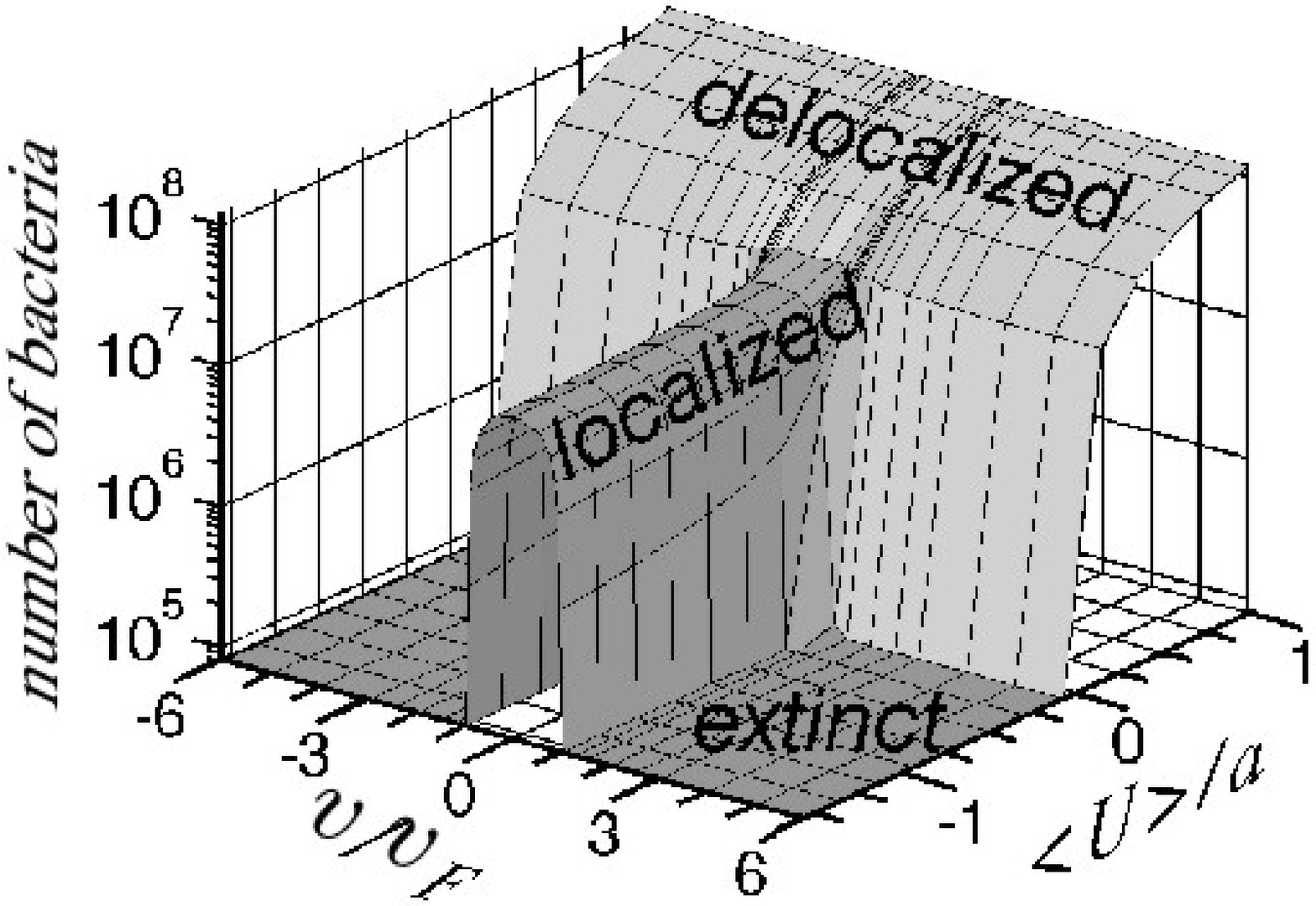}
\vspace{-0.1truein}
\caption{\label{phasedia} Phase diagram computed for the
model, Eq.~\ref{gfisher}, showing the total number of bacteria in
the system as a function of the two control parameters, the
average terrain $\frac{\langle U \rangle}{a}$ and the oasis
velocity $v/v_F$. There are three sustainable states: localized,
where the bacterial colony survives only in the moving oasis;
delocalized, where the bacterial population grows throughout the
system, although the growth is largest inside the oasis region
(see Fig.~\ref{locdeloc}); and extinction, where the bacteria die
out everywhere, even inside the moving oasis. Note that the
diagram is symmetric about $v/v_F = 0$. The parameter values
for the simulation were: $a$ = 4.76 $\times$ 10$^{-4}$ s$^{-1}$,
$b$ = 7.93 $\times$ 10$^{-11}$ cm/s, $D$ = 6 $\times$ 10$^{-5}$
cm$^2$/s, $W$ = 3 cm, $L$ = 20.32 cm, 
and $v_F$ = 3.37 $\times$ 10$^{-4}$ cm/s.}
\end{figure}

The spatiotemporal behavior of the colony under conditions close
to the extinction-localization transition is shown in the middle
column of Fig.~\ref{expsim}. Close to this transition we observe
in the experiment a double-peaked oscillating front, as can be
seen in Fig.~\ref{oscillate}.  These oscillations persist for many
generations. The parameters in the simulations of the model
Eq.~\ref{gfisher} have been varied widely, but no oscillations
have been found.

\noindent {\fntsm Transitions in
bacterial populations.} The one week time required for
measurements for each set of parameter values made detailed
laboratory investigations of the transitions impractical. However,
the numerical simulations of Eq.~\ref{gfisher} were conducted 
for wide ranges in $\frac{\langle U \rangle}{a}$ and
$v/v_F$. The simulations revealed not only the three distinct
states but also information on the transitions among these states:
extinction-localization, extinction-delocalization, and
localization-delocalization, as shown in Fig.~\ref{phasedia}.
The phase diagram is plotted in parameter ranges near 
the transitions, while the experiments were run
at $\frac{\langle U \rangle}{a}$ = -24.5, off the plot.  
Simulations conducted using experimental parameters show 
that the phase diagram for those values can be extrapolated 
from Fig.~\ref{phasedia}.  

\begin{figure}
\centering
\includegraphics[width=4.00truein]{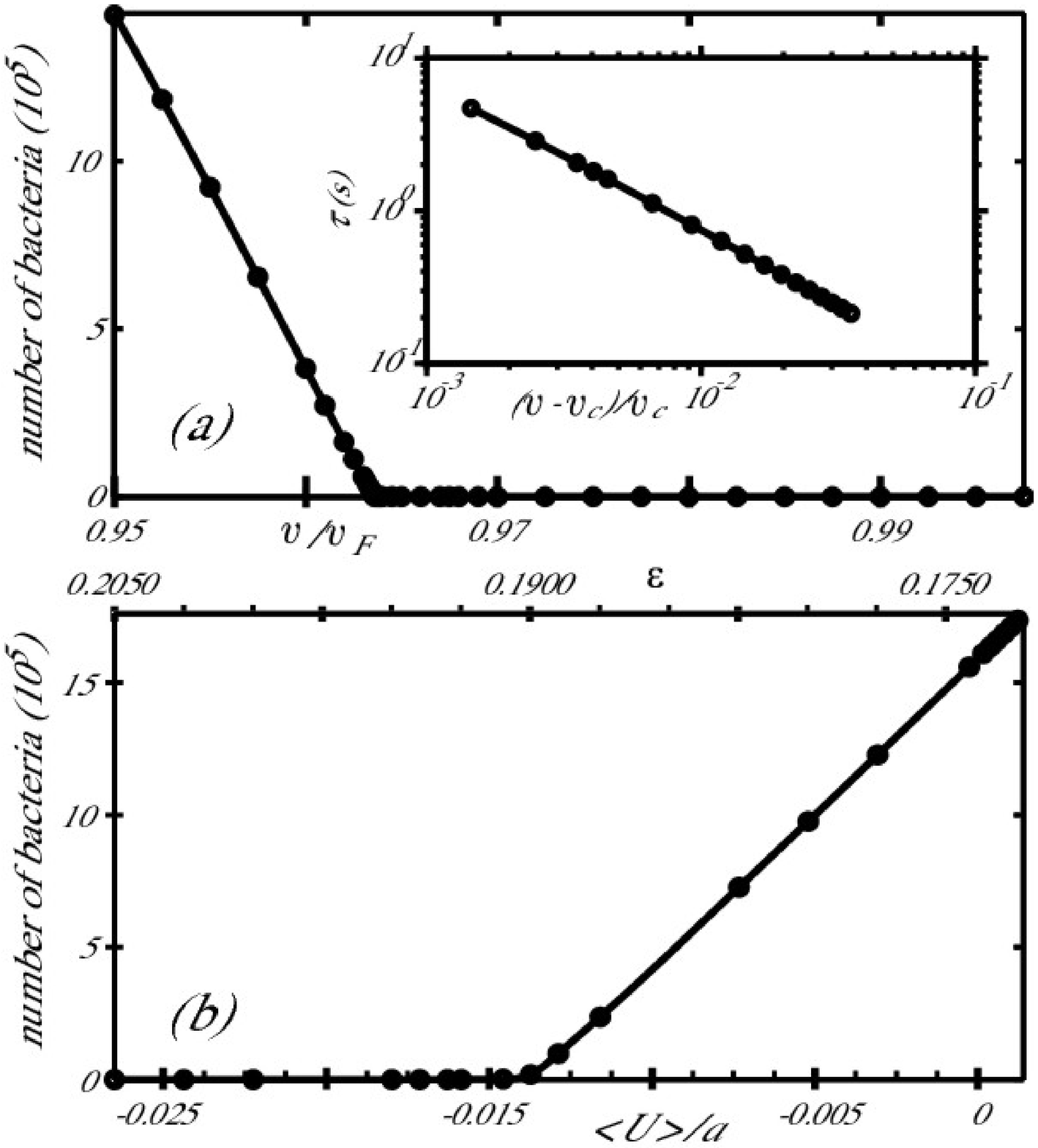}
\vspace{-0.1truein}
\caption{\label{transition} (a) Transition from a
population localized in the oasis to extinction of all bacteria as
the normalized oasis velocity $v/v_F$ was increased
in a simulation of \protect{Eq.~\ref{gfisher}}; the average
terrain was held fixed, $\frac{\langle U \rangle}{a} = -0.49$. The
inset illustrates critical slowing down as the transition at
$v=v_c$ was approached from above: the relaxation time diverges,
$\tau \propto (v-v_c)^{-\gamma}$ with $\gamma=0.98\pm 0.03$. 
Similar results were obtained when approaching the transition
from below. (b) The transition between extinction and
delocalization with fixed $v/v_F$=2, as a function of
$\frac{\langle U \rangle}{a}$ (or, as shown along the top of the
graph, as a function of the attenuation factor $\epsilon$).
Other parameter values are given in Fig.~\ref{phasedia}. }
\end{figure}

Studies of behavior near the transitions were made with $v/v_F$
changed in steps as small as 0.0001 and $\frac{\langle U
\rangle}{a}$ changed in steps as small as 0.0005.
These studies showed that each
transition was continuous, as Fig.~\ref{transition} illustrates.
In addition, scans were made for both increasing and decreasing
values of the control parameters, and the results were found to be
independent of the direction in which the control parameter
changed; this absence of hysteresis provides further evidence that
the transitions are continuous.

To characterize the transitions, we examined the relaxation of the
numerical solution to Eq.~\ref{gfisher} following a change in
control parameter values. In each case, the bacterial
concentration was found to relax exponentially to its final value.
Thus a characteristic
decay time $\tau$ could be determined for each set of control
parameters. These times became very long as a transition was
approached, indicating the critical slowing down that is
characteristic of supercritical bifurcations \cite{CrHo:93}. 
The divergence in relaxation time as the localization-extinction
transition is approached is shown in Fig.~\ref{transition}.

\begin{figure}
\centering
\includegraphics[width=4.000truein,angle=0]{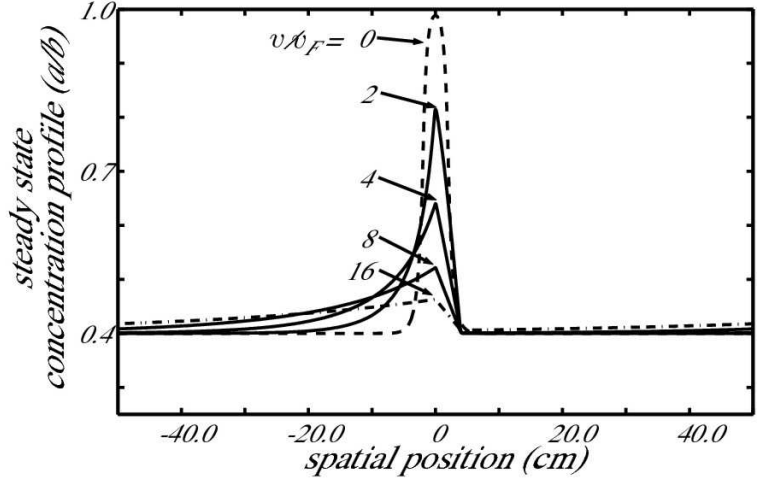}
\vspace{-0.1truein}
\caption{\label{locdeloc} The asymptotic spatial
distribution of an extended (delocalized) 
bacterial population computed in the reference
frame of an oasis moving with different relative velocities
$v/v_F$. The average terrain was held fixed, $\langle U \rangle/a
= 0.424$, $D=1.2\times 10^{-4}$ cm$^2$/s, 
$v_F=4.78\times 10^{-4}$ cm/s, 
$a = 4.76\times10^{-4}$ s$^{-1}$, $\epsilon = -0.4$, $W = 4$ cm,
$L = 1000$ cm, $b = 7.93\times10^{-11}$ cm/s.} 
\end{figure}
For $\frac{\langle U \rangle}{a}$ large enough (i.e., for $a$
and/or $W$ large and/or $\epsilon$ small or negative), 
the average terrain experienced by each 
bacterium was sufficient to support life
everywhere; the population, initially limited to the oasis,
became delocalized. However, the population remained larger in the
oasis region, as Fig.~\ref{locdeloc} illustrates for a range of
increasing oasis velocities.

\vspace{0.1truein}

\noindent
{\fntmed Discussion}\\
A system's response to changes in local and global
conditions is part of the knowledge required for applications
such as achieving precision drug delivery 
e.g, localizing a viral vector in its target cell during
gene therapy.  The moving oasis could correspond 
to a flow of plasma past a 
fixed region of favorable growth conditions.

Our study has concerned bacterial populations that initially grew
in a stationary oasis that lay within a desert; then the oasis was
made to move with a constant velocity. We found that the
population could be sustained or could die out, depending on the
oasis velocity and the average terrain, which depended on
harshness of the desert and the size of the oasis relative to the
size of the desert. 
Results from our numerical simulations of a
model were in qualitative accord with the laboratory observations,
although the model failed to capture the observed double-peaked
oscillations found for the spatial distribution of bacteria in the
moving oasis. Contrary to previous theoretical investigations \cite{DNSh:00},
our simulations show three instead of four sustainable states 
in the range of biologically relevant parameters. 
The simulations of the model yielded deeper insights
into the mechanisms governing the colony behavior than could be
obtained from the experiments alone. In particular, the
simulations showed that the transitions between the different
states were continuous and nonhysteretic. Further, the simulations
yielded a delocalized state, where the bacterial population grew
throughout the entire system; this state was not accessible in the
experiments.

\vspace{0.1truein}

\noindent The authors thank Laura Runyen-Janecky for supplying the
RW 120 $E.\ coli$ bacteria and for her help, and Nicolas Perry for
his help. They also thank Karen Dahmen, David
Nelson, N. M. Shnerb, Andrea Bertozzi and Nitant Kenkre for
helpful discussions. This work was supported by the Engineering
Research Program of the Office of Basic Energy Sciences of DOE and
the Robert A. Welch Foundation (ALL, BAM, HLS), and by NIH and NSF
(GT, BL, JK).

\pagebreak
\bibliography{bacdat}

\end{document}